\documentclass[prl,twocolumn,floatfix,showpacs,tightenlines]{revtex4}
\usepackage{graphicx}
\begin{document}

\title{Nonequilibrium dynamical mean-field theory} 
\author{J.~K.~Freericks$^*$, V.~M.~Turkowski$^*$,
 and V.~Zlati\'c$^\dagger$}
\affiliation{$^*$Department of Physics, Georgetown University, Washington, DC
20057, U.S.A.}
\affiliation{$^\dagger$Institute of Physics, Bijenicka c. 46, P. O. B. 304,
1000 Zagreb, Croatia}

\date{\today}

\begin{abstract}
The many-body formalism for dynamical mean-field theory is extended to treat
nonequilibrium problems.  We illustrate how the formalism works by examining 
the transient decay of the oscillating current that is driven by a large 
electric
field turned on at time $t=0$. We show how the Bloch oscillations are quenched
by the electron-electron interactions, and how their character changes 
dramatically for a Mott insulator.
\end{abstract}


\maketitle

{\it Introduction}. Dynamical mean-field theory (DMFT) was introduced in 1989 as
a technique to solve the quantum many-body problem by taking the limit where
the number of spatial dimensions goes to infinity~\cite{metzner_vollhardt_1989}.
In this limit, with a 
proper scaling of the hopping matrix elements, the electron-electron 
correlations are described by a local self-energy.  Hence the many-body
problem on a lattice is mapped onto an effective many-body problem
for a single-site impurity (in a time-dependent field), with a self-consistency 
condition that fixes the time-dependent field so that the Green's function
for the impurity is identical with the local Green's function for the lattice. 
Since then, DMFT has been employed to solve virtually all many-body problems
described by model Hamiltonians~\cite{kotliar_review}, has been 
generalized to describe strong
electron correlations in real materials~\cite{kotliar_review2} and to
describe inhomogeneous systems~\cite{potthoff_nolting_1999,freericks_book}.  
All of this work has focused on the equilibrium case.
In this contribution, we illustrate how to generalize DMFT to nonequilibrium
situations, and we present results for how the Bloch oscillations of a 
strongly correlated material
are quenched by electron-electron interactions, and how their character changes
after the Mott metal-insulator transition.

Bloch~\cite{bloch_1928} and Zener~\cite{zener_1934} theorized that electrons
on a lattice undergo an oscillatory motion when placed in a uniform
static electric field, because the electron wavevector, which evolves under
the electric field, is Bragg reflected whenever it reaches a Brillouin zone 
boundary. But in metals, Bloch oscillations have never been seen, because
the electron relaxation time is so short, the electrons are scattered 
before they reach the zone boundary and
Bragg reflect.  Bloch oscillations have been observed in semiconducting
heterostructures~\cite{bloch_semi}, Josephson junctions~\cite{bloch_josephson},
and cold-atom systems~\cite{bloch_atom}.  

\begin{figure}[h]
\centering{
\includegraphics[width=1.5in,angle=0]{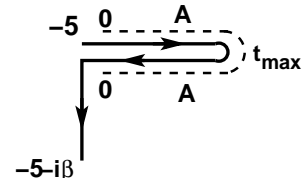}}
\caption{
Kadanoff-Baym-Keldysh contour for the two-time Green's functions
in the nonequilibrium case. We take the contour to run from $-5$ to
$t_{max}$ and back, and then extends downward parallel to the imaginary
axis for a distance of $\beta$.  The field is turned on at $t=0$;
{\it i.e.}, the vector potential is nonzero only for positive times.
}
\label{fig: contour}
\end{figure}

{\it Formalism}. The many-body formalism for nonequilibrium dynamical
mean-field theory is straightforward to develop within the Kadanoff-Baym-Keldysh
approach~\cite{kadanoff_baym_1962,keldysh_1965}.
Because nonequilibrium problems are not time-translation invariant, we need
to work with Green's functions that depend on two times.  There are two
independent Green's functions that need to be determined---the retarded
Green's function $G^R$, which describes the density of available 
quantum-mechanical
states, and the lesser Green's function $G^<$, which determines how electrons 
occupy those quantum states.  Both Green's functions can be extracted from
the contour-ordered Green's function, which is defined for 
any two time values that lie on the Kadanoff-Baym-Keldysh contour shown in
Fig.~\ref{fig: contour}.  We imagine our system to be in equilibrium until
time $t=0$ where a field is turned on.  The contour starts at some time
before the field is turned on, runs out to a maximal time, then returns to
the original time, and finally moves parallel to the negative imaginary
axis a distance $\beta$ (equal to the inverse of the temperature of the
original equilibrium distribution). 

Since the
many-body perturbation theory diagrams are identical in structure
for equilibrium and nonequilibrium perturbation theories~\cite{langreth},
the perturbative analysis of Metzner~\cite{metzner_1991} guarantees that
the nonequilibrium self-energy is also local in DMFT.  Hence, the nonequilibrium
DMFT problem can be mapped onto an impurity problem in time-dependent
fields, just like the equilibrium problem, except now the fields have two
time arguments. The basic structure of the iterative approach to solving the
DMFT equations~\cite{jarrell_1992} continues to hold.  We start with a guess
for the self-energy (which is usually chosen to be equal to the 
equilibrium self-energy), then
we sum the momentum-dependent Green's function over the Brillouin zone to 
produce the local Green's function. Next the dynamical mean-field for
the impurity problem is extracted by using Dyson's equation for the 
local Green's function and self-energy, the impurity problem is solved in the
dynamical mean-field to produce the impurity Green's function, and Dyson's
equation is used again to extract the impurity self-energy.  In the 
self-consistent solution of the DMFT equations, the impurity self-energy will
be equal to the lattice self-energy.  If they are different, then the new
lattice self-energy is taken to be equal to the new impurity self-energy,
and the loop is iterated until it converges.  The nonequilibrium algorithm
is modified as follows: (i) the
summation over the Brillouin zone now requires at least a double integral
over two band energies; (ii) the Green's functions are described by discrete
matrices with time indices that run over the contour; and (iii) the impurity
problem solver must be generalized to the nonequilibrium case.

For concreteness, we assume the electric field ${\bf E}(t)$ is spatially 
uniform, but can depend on time (we assume that the magnetic field, however, is
small, and neglect all magnetic-field effects). We work in the Hamiltonian
gauge, where the scalar potential vanishes, and the electric field is determined
by a time derivative of the vector potential ${\bf E}(t)=-d{\bf A}(t)/dt$,
in units where $c=1$. The noninteracting problem
of Bloch electrons in an electric field can be solved exactly by using the
Peierls substitution~\cite{peierls_1933,turkowski_freericks_2005}, 
and if we take the electric field to lie along the diagonal direction, 
then the noninteracting momentum-dependent Green's functions on the
lattice depend only on two explicit functions of momentum 
\begin{equation}
\epsilon_{\bf k}=-\frac{t^*}{2\sqrt{d}}\sum_{i=1}^d\cos {\bf k}_i,
\quad
\bar \epsilon_{\bf k}=-\frac{t^*}{2\sqrt{d}}\sum_{i=1}^d\sin {\bf k}_i,
\label{eq: band_energies}
\end{equation}
rather than all components of the momentum.
Here we set the lattice constant $a$ equal to 1, and we consider the case
of nearest-neighbor hopping on a hypercubic lattice in $d$-dimensions with
a hopping parameter $t=t^*/2\sqrt{d}$; $t^*$ will be taken as the energy unit.
In the limit $d\rightarrow\infty$, the two ``band energies'' are distributed 
with a joint Gaussian density of states~\cite{schmidt_monien_1999}
\begin{equation}
\rho(\epsilon_{\bf k},\bar\epsilon_{\bf k})
=\frac{1}{\pi}e^{-\epsilon_{\bf k}^2}e^{-\bar\epsilon_{\bf k}^2}.
\label{eq: joint_dos}
\end{equation}
In the interacting case, the dressed contour-ordered
Green's function satisfies Dyson's 
equation, with a local self-energy, so that
\begin{equation}
G({\bf k},t,t^\prime)=\left [ G_0^{-1}({\bf k},t,t^\prime)-
\Sigma(t,t^\prime)\right ]^{-1},
\label{eq: k_dyson}
\end{equation}
where the Green's functions and self-energy are continuous matrix operators
defined on the contour ({\it i. e.}, the time indices of the matrices
run along the contour), and the $-1$ superscripts denote the matrix inverse
of the respective operators. The noninteracting Green's function in a field
(for both times larger than 0; one can easily work out the generalizations for 
cases when the field has not been turned on) is
\begin{eqnarray}
G_0({\bf k},t,t^\prime)&=&i[f(\epsilon_{\bf k}-\mu)-\theta_c(t,t^\prime)]
e^{-i\mu(t-t^\prime)}\label{eq: g0_def}\\
&\times&
e^{-i\epsilon_{\bf k}(\sin eEt-\sin eEt^\prime)/eE}\nonumber\\
&\times&
e^{-i\bar\epsilon_{\bf k}(\cos eEt-
\cos eEt^\prime)/eE},
\nonumber
\end{eqnarray}
where $e$ is the electron charge, $E$ is one component of the electric
field along a Cartesian axis (all components are equal for a field directed
along the diagonal), and $f(x)=1/[1+\exp(x)]$ is the Fermi-Dirac
distribution (we set $\hbar =1$).  The symbol $\theta_c(t,t^\prime)$ is
equal to one if $t$ is ahead of $t^\prime$ on the contour and is zero
otherwise. Calculating the local Green's function requires evaluating
a two-dimensional integral over $\rho(\epsilon_{\bf k},\bar\epsilon_{\bf k})$ 
of a matrix-valued integrand, which requires a matrix inversion to determine it.
Once the local Green's function $G$ has been found, we use Dyson's equation
to extract the dynamical mean-field, denoted $\lambda(t,t^\prime)$,
which satisfies
\begin{equation}
\lambda(t,t^\prime)=(i\partial^c_t+\mu)\delta_c(t,t^\prime)-G^{-1}(t,t^\prime)
-\Sigma(t,t^\prime),
\label{eq: lambda_gen}
\end{equation}
where the derivative with respect to time is taken along the contour, and the
delta function is defined on the contour such that $\int_c dt 
\delta_c(t,t^\prime) F(t)=F(t^\prime)$. 

Next, the impurity problem must be solved for electrons evolving in the
dynamical mean field.  In general, algorithms have not yet been developed to
solve this problem for all Hamiltonians, but the impurity problem can be
immediately solved for the spinless
Falicov-Kimball model~\cite{falicov_kimball_1969}, which involves single-band
conduction
electrons hopping on a lattice, and localized electrons which do not move
but do interact with the conduction electrons when they are in the same
unit cell via a screened Coulomb interaction $U$.  The Hamiltonian (in the
absence of a field) is then
\begin{equation}
\mathcal{H}=-\frac{t^*}{2\sqrt{d}}\sum_{\langle ij\rangle}
(c^\dagger_ic^{}_j+c^\dagger_jc^{}_i)+U\sum_i w_ic^\dagger_ic^{}_i.
\label{eq: hamiltonian}
\end{equation}
Here, we have $c^\dagger_i$ ($c^{}_i$) create (annihilate) a spinless conduction
electron at site $i$ and $w_i=0$ or 1 is the localized electron number operator
at site $i$.  Although the Falicov-Kimball model is a simple many-body physics
model, it does have a Mott metal-insulator transition
(but the model does not include Zener tunneling because there are 
no higher energy bands). The solution to the impurity problem can be found by 
solving the equations of motion for the contour-ordered Green's function
resulting in
\begin{eqnarray}
G_{\rm imp}(t,t^\prime)&=&(1-w_1)
\left [ (i\partial^c_t+\mu)\delta_c(t,t^\prime)-\lambda(t,t^\prime)
\right ]^{-1}\nonumber\\
&+&w_1
\left [ (i\partial^c_t+\mu-U)\delta_c(t,t^\prime)
-\lambda(t,t^\prime)\right ]^{-1}
\label{eq: g_imp_solve}
\end{eqnarray}
with $w_1$ the average localized electron filling.  The Dyson equation in
Eq.~(\ref{eq: lambda_gen}) is then employed to extract the impurity
self-energy, and the algorithm is iterated until it converges.

There are a number of important technical details  
in these calculations.  First, we discretize the contour (we choose a real-time
spacing $\Delta t$ which varies from 0.1 to 0.0333, and we fix the spacing
along the imaginary axis to $\Delta \tau=0.1i$) and evaluate integrals over the 
contour by discrete summations using the midpoint rectangular
integration rule.  The matrix operators are general complex matrices,
which are manipulated using standard linear algebra packages 
(the largest matrices used are about $2200\times 2200$).  
In addition, the delta function changes sign along the negative real time 
branch, and
is imaginary along the last branch of the contour while the derivative of the
delta function is evaluated by a two-point discretization involving the
diagonal and the first subdiagonal, but one also needs to include one
matrix element at the upper right corner to preserve the proper boundary 
conditions. 

We perform the two-dimensional energy integration by Gaussian
quadrature with both 100 and 101 points in each dimension, and we average the
two results.  Since the calculation of each matrix in the integrand of the
integral is independent of every other quadrature point, this part of the
code is easily parallelized.  We require 20,201 matrix inversions
for each DMFT iteration.  The impurity solver
is a serial code, that cannot be parallelized, because the matrix operations
need to all be performed in turn.  We typically require between ten and fifty
iterations to reach convergence of the results (the total computer time for the
calculations presented here was about 600,000 cpu-hours on a Cray XT3).  
Once converged, we calculate the current by evaluating the operator average
\begin{equation}
\langle {\bf j}(t)\rangle=-ei\sum_{\bf k}{\bf v}({\bf k}-eEt)G^<({\bf k},t,t).
\label{eq: current_def}
\end{equation}
The velocity component is $v_i({\bf k})=t^*\sin({\bf k}_i)/2\sqrt{d}$, and
all components of the current are equal when the field lies along the
diagonal.
We also calculate the equal time retarded and lesser Green's functions
and their first two derivatives and compare the results to the exact
values~\cite{turkowski_freericks_2006}.  In general, these ``moments'' are
quite accurate as the step size is made smaller, and we find that if we
use a Lagrange interpolation formula to extrapolate the results to $\Delta t=0$,
we can achieve even higher accuracy for most values of $U$. Details of these
numerical issues and of the accuracies will be presented 
elsewhere~\cite{freericks_turkowski_zlatic_2006}.

\begin{figure}[h]
\centering{
\includegraphics[width=3.3in,angle=0]{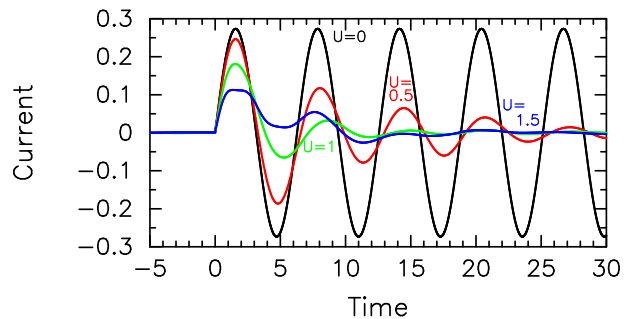}
}
\caption{
Scaled nonequilibrium current for different values of $U$. Note how the
Bloch oscillations are rapidly reduced in amplitude as the scattering
increases. (Color on-line.)
}
\label{fig: current1}
\end{figure}

{\it Numerical results}. We produce numerical calculations of the nonequilibrium
current as a function of time for the case of half-filling, where the conduction
electron and the localized electron fillings are each equal to 0.5.  This system
has a metal-insulator transition at $U=\sqrt{2}$.
In the case where there is no scattering ($U=0$),
the Bloch oscillations continue forever.  In the presence of scattering, the
Bloch oscillations maintain the same approximate ``periodicity'', but the 
amplitude decays.  In Fig.~\ref{fig: current1}, we plot the current for
the noninteracting case, the case of a strongly scattering metal $(U=0.5$, red),
the case of an anomalous metal $(U=1$, green), and of a near critical insulator
$(U=1.5$, blue). The initial temperature of the system satisfied $\beta=10$, and
the field is turned on at $t=0$.  The electric field is set equal to one
in magnitude, $E=1$.  Note how the Bloch oscillations are damped as the 
scattering increases.  Although a Boltmann equation approach always predicts
that the oscillations are damped and disappear on a time-scale on the order
of the relaxation time, and approach a constant steady state, we do not
see this full evolution within the time-window that we performed these
calculations.  Most of the data given here involve a scaling of the 
data with $\Delta t=0.1$, 0.067, 0.05, and 0.04 to the $\Delta t\rightarrow 0$
limit.  Note that the data for a fixed step size in time always shows a small
current for $t<0$, but when scaled, the current is completely flat and
vanishes for negative times (we estimate the scaled data has a relative error
of less than 1\%).  The Bloch oscillations are always nearly as
large as in the noninteracting case, and then they begin to decay.  We are
unable to determine whether they decay to a constant value as predicted by the
Boltzmann equation, or whether there are oscillations present in the steady
state. In the quantum-mechanical system, there are two relevant time scales,
the average time, and the relative time.  The relative time governs the decay
of the quasiparticle-like excitations, and this decay becomes rapid as the 
scattering increases.  The average time governs the Bloch oscillations,
and it is not obvious from either the formalism or our results whether 
the steady-state current must be a constant, or whether it can oscillate
if the electric field is large enough (of course, with a period as short
as these Bloch oscillations would have, they could not be measured 
in an experiment).

\begin{figure}[h]
\centering{
\includegraphics[width=3.1in,angle=0]{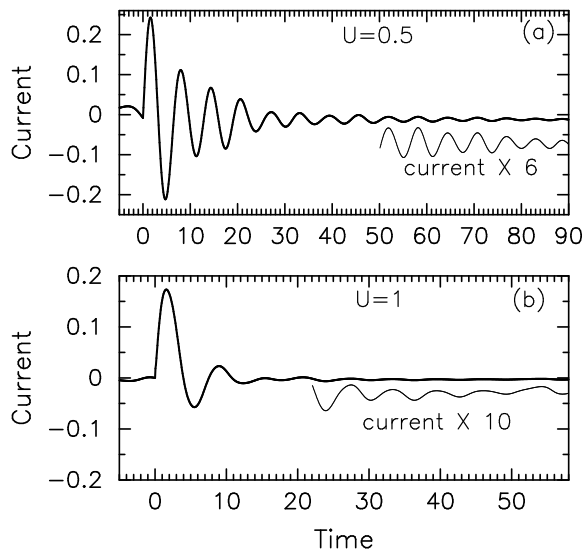}
}
\caption{
Nonequilibrium current for  (a) $U=0.5$  ($\Delta t=0.1$) 
and (b) $U=1$  (scaled from $\Delta t=0.1$ and $0.0667$) and longer times. 
Note how the Bloch oscillations are still present but become more erratic
at large times. The current is multiplied by either 6 or 10 to enhance it
at large times.
}
\label{fig: current2}
\end{figure}

\begin{figure}[h]
\centering{
\includegraphics[width=3.1in,angle=0]{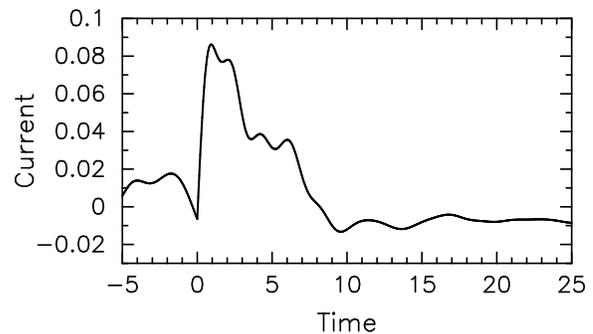}
}
\caption{
Nonequilibrium current for $U=2$ and $\Delta t=0.0333$. Note how the
Bloch oscillations are now quite irregular and how there is substantial
current before $t=0$, because the data is not scaled.
}
\label{fig: current3}
\end{figure}

In Fig.~\ref{fig: current2}, we show results for the current with $U=0.5$ and
$U=1$.  We take the time 
window to be larger here to see if we can shed any further light on how the
data evolves to the steady state but the time window is still too short.
In Fig.~\ref{fig: current3}, we plot the current as
a function of time for the small-gap Mott insulator with $U=2$.  It is much
harder to achieve convergence for these results, and the scaling approach
does not seem to work well, as the calculated moments are less accurate
for the scaled data, than for the data at the smallest $\Delta t$ value
(which is 0.0333 for the $U=2$ data; relative errors here are probably at 
the 10\% level).  Note how there is nonzero 
current at negative times, indicating that this data is not quite as accurate
as the data with smaller $U$ values.  Note further, that the oscillatory 
behavior is quite irregular here, and it is not 
an exponentially decaying Bloch oscillation anymore.  The Bloch oscillations
rapidly change their character as the metal-insulator transition is crossed.

{\it Summary}.  We have developed the formalism for nonequilibrium
dynamical mean-field theory.  The basic method is similar to that of the
equilibrium case, except we need to work in a real-time representation for
all Green's functions, self-energies, and dynamical mean fields. The
summation over momentum to yield the local Green's function now involves a
two-dimensional integration of a matrix-valued integrand.
We presented numerical results for the
Bloch oscillations in the presence of a large electric field, and showed
how they decay as a function of time when there is electron-electron
scattering.  In the transient response calculations presented here, we
cannot determine whether oscillations are present in the steady state.  We 
are currently working on
a steady-state formalism should be able to shed light on that issue.  

{\it Acknowledgments}.  We acknowledge useful conversations with
J. Serene and A. Joura.  This work is supported by the N. S. F. under grant
number DMR-0210717 and by the O. N. R. under grant number N000140510078.
Supercomputer time was provided by the ERDC XT3 under a CAP phase II 
project in the winter of 2006.

\end{document}